\documentclass[12pt]{iopart}
\usepackage{graphicx}
\usepackage{iopams}

\begin{document}

\title[Spin currents in a normal two-dimensional electron gas in contact ...]{Spin currents in a normal two-dimensional electron gas in contact with a spin-orbit interaction region}

\author{Aleksei\,A\,Sukhanov, Vladimir\,A\,Sablikov and Yurii\,Ya\,Tkach}

\address{Kotel'nikov Institute of Radio Engineering and Electronics,
Russian Academy of Sciences, Fryazino, Moscow District, 141190, Russia}

%which is laterally contacted
\begin{abstract}
Spin effects in a normal two-dimensional (2D) electron gas in lateral contact with a 2D region with spin-orbit interaction are studied. The peculiarity of this system is the presence of spin-dependent scattering of electrons from the interface. This results in an equilibrium edge spin current and nontrivial spin responses to a particle current. We investigate the spatial distribution of the spin currents and spin density under non-equilibrium conditions caused by a ballistic electron current flowing normal or parallel to the interface. The parallel electron current is found to generate the spin density near the interface and to change the edge spin current. The perpendicular electron current changes the edge spin current proportionally to the electron current and produces a bulk spin current penetrating deep into the normal region. This spin current has two components, one of which is directed normal to the interface and polarized parallel to it, and the second is parallel to the interface and is polarized in the plane perpendicular to the contact line. Both spin currents have a high degree of polarization ($\sim 40-60\%$).
\end{abstract}
%\PACS{\textbf{73.20.-r, 72.25.Hg, 73.43.Jn, 85.75.-d}}
% Submitted to: J. Phys.C: Solid State Phys.
\maketitle

%\begin{twocolumn}

\section{Introduction }

Spin transport in semiconductor nanostructures with spin-orbit interaction (SOI) attracts much attention, stimulated by the interesting spin dynamics in systems with non-conserving spin and possible applications in creating and manipulating spin polarization in nonmagnetic systems by purely electric methods using SOI~\cite{Awsch,Zutic, Fabian,Awschalom}. Spin currents are widely studied in the cases where there is a spin Hall effect, either intrinsic~\cite{Mur,Sino,Kato,Wund,Engel} or extrinsic~\cite{Dyak,Hirsch,Kato}. The basic experimentally observable effect is the spin accumulation near lateral edges of the sample if the electric current is driven along it. The spin density is accumulated because the transverse spin current does not pass through the edges.

 A somewhat different effect appears due to electron scattering on the sample boundary if the particle current flows parallel to it. In this case a spin polarization is formed near the boundary and the spin density distribution has a rather complicated form~\cite{Usaj,Rey,Xing,Silvestrov}.

In this paper another system is studied, in which electrons can pass through the boundary of the SOI region. We consider a 2D system composed of an electron gas with SOI and normal (N) 2D electron gas without SOI, lying in one plane. This system is designated as SOI/N contact. The main purpose is to find out whether the spin currents, which exist in the SOI region due to the coupling between spin and orbital degrees of freedom, penetrate into the N region and how the spin density is distributed there under the non-equilibrium conditions when an electron current flows.

This system is also interesting because it reveals a new aspect of the problem of equilibrium spin currents. The existence of background spin currents in thermodynamic equilibrium was pointed out by Rashba~\cite{Rashba} for infinite 2D electron gas. Usaj et al~\cite{Usaj} showed that equilibrium spin current exists also in a bounded 2D electron system with SOI, the spin current flowing near the edges of the sample.

The phenomenon of equilibrium spin currents has provoked an intense discussion on the discerning of spin currents and their definition. The question arose since spin currents in the SOI medium are not unambiguously defined ~\cite{Rashba,Son,Sun}. Sun et al~\cite{Sun} found that the equilibrium spin current exists in a quantum ring even if the SOI is absent in some segment.  In this connection, of more interest is a system, in which 2D SOI medium is in contact with a bulky normal 2D electron gas where the spin current is well defined. The question is whether the equilibrium spin current exceeds the boundary of the SOI medium, entering into the N electron gas and what spin effects appear there.

This question was raised in our previous papers~\cite{Sablikov,Tkach}. We have established that the equilibrium spin current does not pass through the boundary, but an equilibrium edge spin current flows near the boundary in both SOI and N regions. Its density considerably exceeds the density of the bulk equilibrium spin current. An important point is that the spin current is well defined in the N electron gas and its definition is beyond doubt. In the present work we study spin effects resulting from the interface scattering under non-equilibrium conditions appearing when an electron current flows in the system. This question is important since only non-equilibrium transport produces a spin accumulation that can be detected experimentally~\cite{Kato,Crooker}.

The calculations are carried out for a weak deviation from the equilibrium caused by a ballistic electron current flowing normal or parallel to the interface. The electron current is supposed to be created by a difference of chemical potentials $\Delta \mu$ of electrons moving in opposite directions. Two main effects are found. The electron current parallel to interface leads to the spin accumulation near the interface and changes in the edge spin current. The accumulated spin density is proportional to $\Delta \mu$, while the augmentation of the edge spin current is proportional to $\Delta \mu^2$. This means that the non-equilibrium spin current does not depend on the polarity of the electron current. If the electron current is perpendicular to the interface, the edge spin currents vary linearly with $\Delta \mu$. However, the main effect consists in an appearance of the bulk spin current in N region. It has two components. One component is directed parallel to the interface and polarized in the plane perpendicular to it. The second component is directed normal to the boundary and polarized parallel to it. As a matter of fact, it is a spin Hall current caused by the SOI/N interface scattering. Both these spin currents penetrate deep into the normal region and have considerable polarization.

\section{The model}

\begin{figure}
 \centerline{\includegraphics[width=8cm]{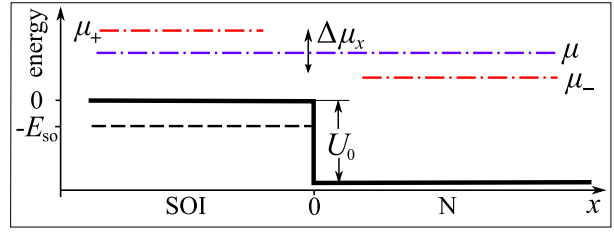}}
\caption{Energy diagram of the SOI/N contact. $U_0$ is the potential step at the interface; $\mu$ is the equilibrium chemical potential; $\Delta \mu_x=\mu_+-\mu_-$ is the difference of chemical potentials of the right- and left- moving electrons in the presence of a perpendicular electron current.}
\label{diagram}
\end{figure}

Consider a 2D SOI/N contact. The energy diagram is shown in Fig.~\ref{diagram}. The SOI and N regions are located correspondingly at $x<0$ and $x>0$. To be specific we suppose that the potential of the SOI region is higher than that of the N region. This arrangement allows one to consider an interesting case where free electrons are present only in the N region, though they penetrate into the SOI barrier. We assume also that the external electric field is too weak to change the equilibrium electron states and the electron current is caused by a non-equilibrium occupation of unperturbed electron states. Such a situation can be realized, for example, if electrons move ballistically and their distribution function is formed by adiabatic contacts, with the applied voltage dropping across the near-contact regions.

The wave functions are determined by the Hamiltonian:
\begin{equation}
H=\frac{p_{x}^{2}+p_{y}^{2}}{2m}+\frac{\alpha}{\hbar}(p_{y}\sigma_{x}-p_{x}\sigma_{y})+U(x),
\label{H}
\end{equation}
where $p_{x,y}$ are electron momentum components; $m$ is the effective mass, which is supposed to be constant all over the structure; $\sigma_{x,y}$ are the Pauli matrices; $\alpha$ is the parameter of SOI ($\alpha=0$ for $x>0$ and $\alpha=\mathrm {const}$ for $x<0$); $U(x)$ is the potential energy, which is also a step function: $U(x)=-U_0\Theta(x)$.

The wave functions of electrons moving to the right (R) and to the left (L) are~\cite{Sablikov}:
\begin{equation}
\varphi_{\lambda,\vec{k}}^{(R)}(x)=\left\{\begin{array}{lc}
|\lambda,k_x,k_y\rangle+\sum\limits_{\lambda'} r_{\lambda,\lambda'}^{R}|\lambda',-k_x,k_y\rangle, &  x<0 \\
\sum\limits_{\lambda'}t_{\lambda,\lambda'}^{R}|\lambda',q_x,k_y\rangle, & x>0 \,,
\end{array}\right.
\label{R}
\end{equation}
\begin{equation}
\varphi_{\lambda,\vec{k}}^{(L)}(x)=\left\{
\begin{array}{lr}|\lambda,-q_x,k_y\rangle+\sum\limits_{\lambda'} r_{\lambda,\lambda'}^{L}|\lambda',q_x,k_y\rangle,&  x>0 \\
\sum\limits_{\lambda'}t_{\lambda,\lambda'}^{L}|\lambda',-k_x,k_y\rangle,& x<0 \,.
\end{array}\right.
\label{L}
\end{equation}
Here $k_{x,y}$ and $q_x$ are components of the electron wave vectors in the SOI and N regions, the index $\lambda$ numbers the branches of the energy spectrum and spin states.

The reflection and transmission matrices, ${r_{\lambda,\lambda'}^{R,L}}$ and ${t_{\lambda,\lambda'}^{R,L}}$, are found using the following boundary conditions~\cite{Sablikov2}:
\begin{equation}
\varphi \Big|_{-0}^{+0}=0\,, \hspace{0,4cm}
\left.\frac{\partial\varphi}{\partial x}\right|_{+0}=\left[\frac{\partial\varphi}{\partial x}-(ik_{so}\sigma_y-\beta k_y\sigma_z)\varphi \right]_{-0}\,.
\label{bound_cond}
\end{equation}
where $k_{so} = m\alpha/\hbar^2$ is the characteristic wave vector of the SOI. The parameter $\beta$ is introduced to take into account the SOI resulting from a potential gradient at the interface. This parameter is usually small. Its value is estimated as~\cite{Sablikov2,Tkach} $\beta=2k_{so}U/(eF_{z})$, with $F_{z}$ being the electric field normal to the 2D layer. For a typical experimental situation of InGaAs quantum wells~\cite{Nitta} with $k_{so}=10^{5}$~cm$^{-1}$, $U=5\cdot 10^{-2}$~eV$ $ and $F_{z}=10^6$~Vcm$^{-1}$ one finds $\beta=0.01$.

The basic states $|\lambda,k_x,k_y\rangle$ are characterized by the energy $E$, the wave vector component $k_{y}$, which is invariant since the structure is uniform along the $y$ axis, and the spin index $\lambda$. These states were in detail described previously~\cite{Sablikov,Sablikov2}. Therefore, we only mention that there are three types of states. For $E<-E_{so}$, where $E_{so}=m\alpha^2/(2\hbar^2)$ is the characteristic energy of SOI, the electron states are evanescent. They are characterized by the complex wave vector $k_x$. For $E>-E_{so}$ all states are either propagating, with real $k_x$, or entirely decaying, with imaginary $k_x$. Each of these states exists in a corresponding interval of the tangential momentum $k_{y}$. If $E>0$, the index $\lambda=\pm$ stands for the chirality. In an interval $-E_{so}<E<0$, the electron states have only negative chirality, but there are two kind of states with a given energy. They differ in the wave vector $k_x$ and for convenience are also labeled by the same index $\lambda=\pm$. Of course, all these states are used in calculating the spin density and the currents.

Using the wave functions (\ref{R}) and (\ref{L}), we find the spin current density $\mathcal{I}_i^{\gamma}$ and the spin density $S^{\gamma}$ for a state $|\varphi_{\lambda,\vec{k}}^{(r)}\rangle$, with $r=(R,L)$ denoting the right- or left- moving states:
\begin{equation}
\mathcal{I}_i^{\gamma}(r,\lambda,\vec{k};x)=\frac{\hbar}{4}\left\langle\varphi_{\lambda,\vec{k}}^{(r)}\biggl|\sigma_{\gamma}v_i+v_i\sigma_{\gamma}\biggr|\varphi_{\lambda,\vec{k}}^{(r)}\right\rangle,
\label{SpinCurrent}
\end{equation}
\begin{equation}
S^{\gamma}(r,\lambda,\vec{k};x)=\frac{\hbar}{2}\left\langle\varphi_{\lambda,\vec{k}}^{(r)}\biggl|\sigma_{\gamma}\biggr|\varphi_{\lambda,\vec{k}}^{(r)}\right\rangle,
\label{Spin}
\end{equation}
where $\gamma$ labels the spin components, $\vec{v}$ is the velocity.

The total spin current $\mathcal{J}$ and spin density $\mathcal{S}$ are found by summing equations ~(\ref{SpinCurrent}) and (\ref{Spin}) over all states occupied according to the distribution function.

In the presence of a charge current $J$, electrons have an asymmetric momentum distribution function $f_{\lambda}(\vec{k})$. We simulate this asymmetry by two semicircles in $\vec k$-space with different Fermi momenta and, accordingly, with different chemical potentials $\mu_+$ and $\mu_-$ for electrons moving in opposite directions. This distribution function can be realized when electrons move ballistically between two contacts with the potential difference $\Delta\mu=\mu_+-\mu_-$. In what follows we consider two cases where the electron current is normal or parallel to the boundary. Accordingly, all quantities ($\Delta\mu_j$, $\mathcal{S}_j$ and $\mathcal{J}_j$) are labeled by index $j=(x,y)$ specifying the direction of the electron current.

As a result we obtain the following expressions for the spin current and spin density:
\begin{eqnarray}
\mathcal{J}_{i,j}^{\gamma}(\Delta\mu_j,x)=&\sum_r^{R,L}\sum_{\lambda}^{\pm}\!\!\int_{-U_0}^{\infty}\!\!dE\int\!\! dk_y\times\nonumber \\   &\times  f_{\lambda}(\vec{k},\Delta\mu_j)D_{\lambda}^{(r)}(E,k_{y})\mathcal{I}_{i}^{\gamma}(r,\lambda,\vec{k})\,,
\label{spin_current_x}
\end{eqnarray}
\begin{eqnarray}
\mathcal{S}_j^{\gamma}(\Delta\mu_j,x)=&\sum_{r}^{R,L}\sum_{\lambda}^{\pm}\!\!\int_{-U_{0}}^{\infty}\!\!dE\int \!\! dk_y\times\nonumber \\ &\times f_{\lambda}(\vec{k},\Delta\mu_j)D_{\lambda}^{(r)}(E,k_y)S^{\gamma}(r,\lambda,\vec{k}),
\label{spin_x}
\end{eqnarray}
with $D_{\lambda}^{(r)}\!(E,k_y)$ the density of states with chirality $\lambda$.

For small $\Delta\mu_j\ll E_{so},U_0$ the spin current and density can be expanded in $\Delta\mu_j$:
\begin{eqnarray}
\mathcal{J}_{i,j}^{\gamma}(x)=&\mathcal{J}_i^{\gamma(0)}(1-\delta_{\gamma i})(1-\delta_{i x}\delta_{\gamma z})+\nonumber\\ &+\mathcal{G}_{i,j}^{\gamma}(1-\delta_{\gamma i}) \Delta\mu_{j}+\mathcal{G}_{i,j}^{\gamma (2)}\Delta\mu_j^2\,,
\label{SC_expansion}
\end{eqnarray}

\begin{equation}
\mathcal{S}_j^{\gamma}(x)=\chi_j^{\gamma}(1-\delta_{\gamma j})(1-\delta_{j x}\delta_{\gamma z})\Delta\mu_j.
\label{S_expansion}
\end{equation}
Here the expansion of $\mathcal{J}$ includes the equilibrium spin current $\mathcal{J}_i^{\gamma(0)}$. A quadratic term is also important because some linear terms are equal to zero. To be exact, the linear response of the tangential spin current to the parallel current of particles is absent.

In section 3 we study in detail the response functions of the spin density, $\chi_{j}^{\gamma}$, and the spin current, $\mathcal{G}_{ij}^{\gamma}$. They are functions of the coordinate $x$. The calculations are carried out using the following distribution function:
\begin{equation}
f_{\lambda}(\vec{k},\Delta\mu_j)=\left\{
\begin{array}{lr}
 \Theta(\mu+\Delta\mu_j/2-E), & k_j>0\,,\\
 \Theta(\mu-\Delta\mu_{j}/2-E), & k_j<0\,.
\end{array}
\right.
\end{equation}

Thus, the calculation scheme is the following. First, the matrices of reflection and transmission coefficients, ${r_{\lambda,\lambda'}^{R,L}}$ and ${t_{\lambda,\lambda'}^{R,L}}$, are found from equations (\ref{R})\,-\,(\ref{bound_cond}). Then the partial spin current  and spin density components, $\mathcal{I}_i^{\gamma}$ and $S^{\gamma}$, are determined using equations (\ref{SpinCurrent}) and (\ref{Spin}). Finally, the total spin current and spin density, $\mathcal{J}_{i,j}^{\gamma}(x)$ and $\mathcal{S}_j^{\gamma}(x)$, as functions of $x$ are obtained from equations (\ref{spin_current_x}) and (\ref{spin_x}) using numerical integration.

\section{Results and discussion}

Below, the results of calculations of the spin density and spin currents are presented for two directions of the electron current: normal and parallel to the SOI/N interface.

But we begin with recalling the main results concerning the equilibrium spin current.

\begin{figure}
 \centerline{\includegraphics[width=8cm]{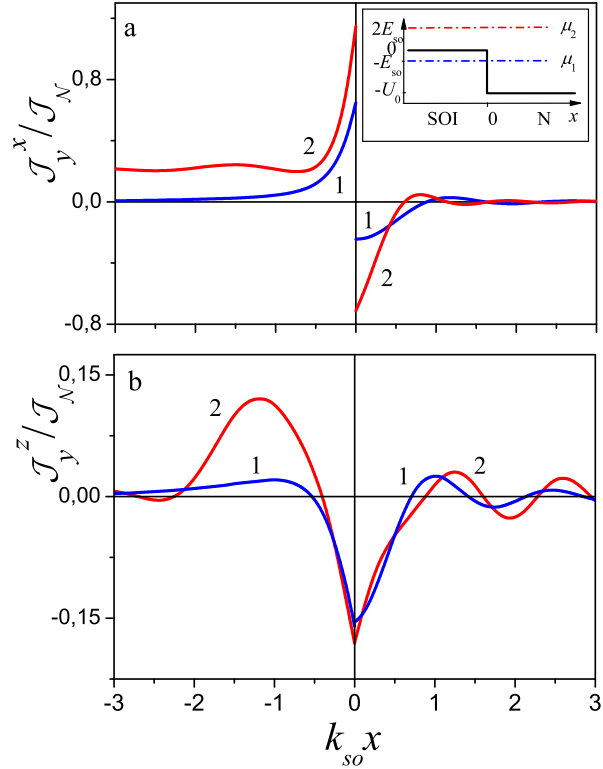}}
 \caption{Spatial distributions of the equilibrium spin current densities $\mathcal{J}_y^{x(0)}(x)$ and $\mathcal{J}_y^{z(0)}$. Curves 1 and 2 correspond to chemical potentials $\mu_{1}$ and $\mu_{2}$, shown in the inset. Calculations were performed for $U_0=6E_{so}$ and $\beta=0.01$.}
\label{equilibrium_SC}
\end{figure}

In unrestricted 2D electron gas with SOI the equilibrium spin current has two nonzero components: $\mathcal{J}_y^{x(0)}=-\mathcal{J}_x^{y(0)}=(2/3\pi) \mathcal{J}_N$, where $\mathcal{J}_N\equiv\hbar^2 k_{so}^3/4m$ ~\cite{Rashba}. Near the SOI/N interface the perpendicular component of the spin current vanishes and it is absent in the N region: $\mathcal{J}_x^{y(0)}(x\geq0)=0$. However, near to the interface there is an edge equilibrium spin current, which flows parallel to the boundary in both regions~\cite{Sablikov}. The density of this current is considerably higher than the bulk density $(2/3\pi)\mathcal{J}_N$, since the edge current is proportional to the SOI constant $\alpha$, while the bulk spin current is proportional to $\alpha^3$. The spin of the edge spin current is polarized in a plane perpendicular to the interface and turns in this plane with increasing distance $x$. Therefore, only two components of the spin current, $\mathcal{J}_y^{x(0)}(x)$ and $\mathcal{J}_y^{z(0)}(x)$, are presented in Fig~\ref{equilibrium_SC}.

The signs of the spin currents $\mathcal{J}_y^x(x)$ in SOI and N regions are seen to be opposite, but the total current is nonzero. It is interesting that the equilibrium edge spin current exists at $\mu<-E_{so}$, i.e. even if free electrons are absent in the bulk of the SOI region. The current is caused by electrons penetrating under the barrier into the SOI region. On increasing the Fermi level above the conduction band bottom in the SOI region, $\mu>-E_{so}$, the spin current appears in the bulk of the SOI region. It increases with $\mu$ to reach a limiting value $(2/3\pi)\mathcal{J}_N$ at $\mu=0$ and then does not change. In contrast, the edge spin current  increases monotonically with $\mu$, approximately as $(\mu+U_{0})^{1/2}$.

Note that the parameter $\beta$ does not affect the results very greatly. Thus, with varying $\beta$ the spin current changes by an amount of the order of $\beta$.

\subsection{Parallel electron current}

If the electron current flows parallel to the SOI/N interface, the picture of the distribution of edge spin currents changes quite a bit. The point is that in this case the linear response of the spin current is absent, $\mathcal{G}_{y}^{x}=\mathcal{G}_{y}^{z}=0$. This occurs for the following reason. The edge spin current is created equally by electrons moving up and down the $y$ axis. If the distribution function $f(k_{y})$ is symmetric, the charge currents of the up- and down-moving electrons cancel each other, while the spin currents, in contrast, are summed. Therefore, in the thermodynamic equilibrium the particle current is absent but the edge spin current is nonzero. A small asymmetric variation of the distribution function $\Delta f(k_y)$ results in the parallel particle current (linear in $\Delta f(k_y)$), but in the first approximation does not change the spin current. The variation of the edge spin current is quadratic in $\Delta f(k_y)$. Hence it is proportional to the square of the electron current and in this approximation does not depend on its polarity.

The augmentation of the spin current, caused by the parallel electron current, is described by the factors $\mathcal{G}_{y,y}^{x(2)}$ and $\mathcal{G}_{y,y}^{z(2)}$  in equation ~(\ref{SC_expansion}). The calculation shows that these factors decrease with distance from the interface and oscillate with the period $\sim\pi/k_{so}$.

\begin{figure}
 \centerline{\includegraphics[width=8cm]{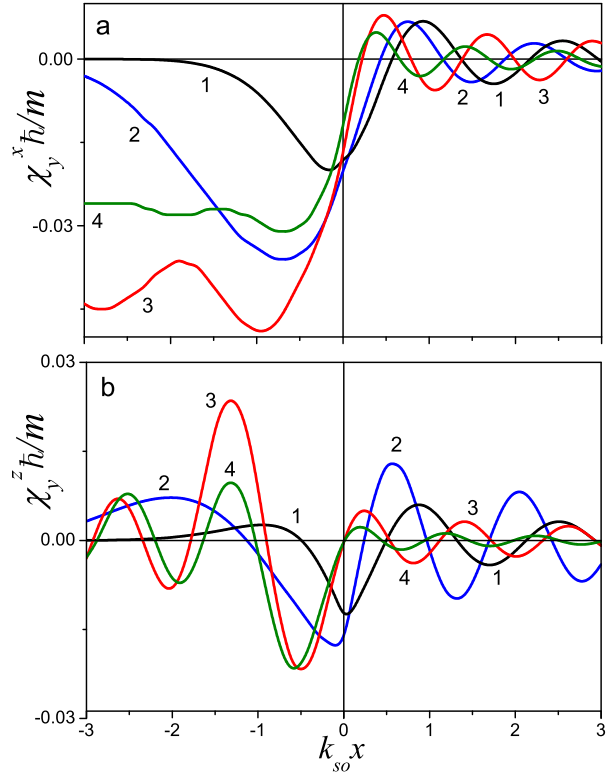}}
\caption{Spatial distribution of the spin susceptibilities $\chi_y^x$ and $\chi_y^z$. Curves 1, 2, 3, 4 correspond to $\mu/E_{so}=-2, -1.1, 0.5, 2.5$. $U_0=6E_{so}$ and $\beta=0.01$.}
\label{spin_dens_y}
\end{figure}

Another stronger effect consists in the spin accumulation near the boundary in both SOI and N regions. The spin density has two nonzero components, $S_y^x$ and $S_y^z$, which are proportional to $\Delta\mu_y$ in the first approximation. The spatial distribution of the spin susceptibility components $\chi_y^x$ and $\chi_y^z$ is presented in Fig.~\ref{spin_dens_y} for different positions of the Fermi level.

The spin susceptibility $\chi_y^x(x)$ is seen to oscillate and fall to zero with the distance $x$ in the N region for all Fermi levels. In the SOI region, $\chi_y^x(x)$ also drops to zero at $-U_{0} <\mu <-E_{so}$. However, at $\mu>-E_{so}$ the spin density in the bulk reaches asymptotically a constant value, which is determined by the difference in occupation between electron states with velocities directed along  $y$ axis and opposite to it.

The spin susceptibility component $\chi_y^z (x)$ oscillates with a period $\sim \pi/k_{so}$ and slowly goes to zero with the distance in the bulk of both SOI and N regions. As a function of the Fermi energy at a given $\Delta\mu_y$, the spin susceptibilities $\chi_y^x$ and $\chi_y^z$ increase in the range $\mu<-E_{so}$ and decrease at $\mu >-E_{so}$.

Numerical estimation of the spin density accumulated near the boundary in InGaAs quantum well for $U_{0}=6E_{so}$, $\mu=E_{so}$ and the electric current density 1~mA/cm give a value of the order of $2\mathcal{S}/\hbar \sim 1\mu m^{-2}$. Recent experiments have shown that such a value of spin density can be detected by Kerr rotation microscopy~\cite{Kato,Crooker}.

\subsection{Normal electron current}

The electron current can flow perpendicularly through the interface only if $\mu>-E_{so}$, i.e. when free electrons are present in the bulk of the SOI region. First, note that in the bulk of the SOI region there is an equilibrium spin current $\mathcal{J}_x^{y(0)}$~\cite{Rashba} flowing normal to the interface, but this current does not pass into the N region~\cite{Sablikov}. In the presence of a non-equilibrium electron current, this spin current keeps unchanged if $\mu>0$, and slightly diminishes proportionally to $[eV/(\mu+E_{so})]^2\ll 1$ otherwise. Thus spin effects in the N region originate solely from the spin-dependent scattering of electrons from the interface. The main effect is the appearance of the spin current in the N region. There are two components of the non-equilibrium spin current: normal and tangential.

\begin{figure}
 \centerline{\includegraphics[width=8cm]{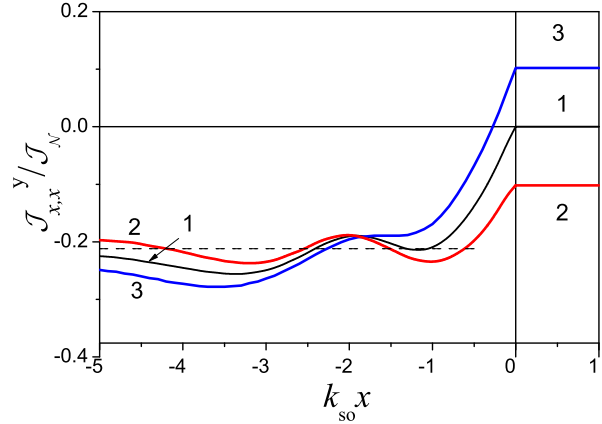}}
\caption{Spatial dependence of the normal components of the spin current $J_{xx}^y$ for different polarities of the electron current. Curve 1 is the equilibrium spin current ($\Delta\mu_x=0$); curve 2 is the spin current in the presence of the electron current flowing from N to SOI region ($\Delta\mu_x=2 E_{so}$); curve 3 is spin current for the electron current of opposed direction ($\Delta\mu_x=-2 E_{so}$). Dashed line indicates the asymptotic value of the spin current in the bulk. $\mu=2E_{so}$, $U_0=6E_{so}$ and $\beta=0.01$.}
\label{SC_normal}
\end{figure}

The spin current component directed normally and polarized in $y$ direction, $\mathcal{J}_{x,x}^y$, arises in both SOI and N regions as a result of the interface scattering. This is a longitudinal spin current effect. Its dependence on the distance $x$ is shown in Fig.~\ref{SC_normal} for two opposed directions of the electron current ($ \Delta\mu> 0$ and $ \Delta\mu <0$). In the SOI region the spatial dependence of $\mathcal{J}_{x,x}^y$ reflects the interference pattern of the incident and reflected electron flows. In contrast, in the N region the spin current does not depend on the distance, since the spin is a conserved quantity there. The spin density is constant as well. The spin polarization (defined as $2\mathcal{J}_{x,x}^y/(\hbar J)$) depends on the chemical potential reaching a maximum (about 50\%) when $\mu$ lies close to the conduction band bottom in the SOI region.

\begin{figure}
 \centerline{\includegraphics[width=8cm]{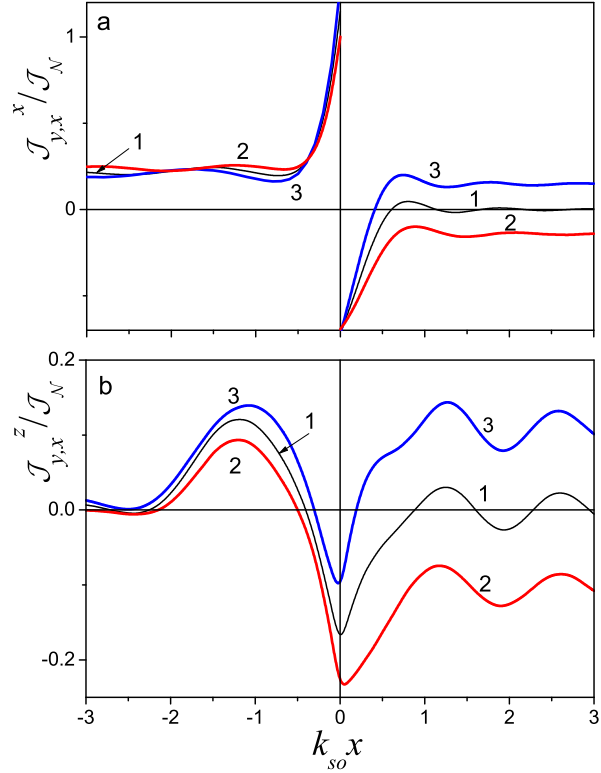}}
 \caption{Distance distribution of the tangential spin currents with spin polarized in $x$ and $z$ directions (panels (a) and (b)), for  different polarities of the electron current. Curves 1 depict the equilibrium current at $\Delta\mu=0$; curves 2 and 3 are the spin currents for opposed directions of the electron current, $\Delta\mu=\pm 2 E_{so}$. $U_0=6E_{so}$ and $\beta=0.01$.}
\label{SC_tang}
\end{figure}

The tangential spin-current has two components of spin polarization lying in the plane perpendicular to the interface, $\mathcal{J}_{y,x}^x$ and $\mathcal{J}_{y,x}^z$. This is a transverse spin current effect. The dependence of the transversal currents on the distance is shown in Fig.~\ref{SC_tang} for opposed polarities of the electron current. Changes of the transverse spin current in the N region under the action of an electron current $J_x$ are, of course, a manifestation of the spin Hall effect caused by the spin-dependent scattering from the interface.

The mechanism of the non-equilibrium spin effects in the bulk of the N region can be understood in a way similar to that suggested in Ref.~\cite{Sablikov} to explain the equilibrium edge spin currents. Consider an unpolarized electron flow falling on the interface from the N region. During the scattering the electrons partially penetrate into the SOI region where an effective magnetic field forces their spin to precess. In the case of the Rashba SOI, this field is perpendicular to the wave vector and lies in the plane. As a result of the spin precession, the spin polarization of reflected electrons changes. Spin components $S_x$ and $S_z$ of electrons with equal $k_x$ but opposite $k_y$ turn out to be of opposite sign and cancel each other. Therefore under the equilibrium conditions the spin density is absent, but the edge spin current exists since the partial spin currents of electrons with opposed $k_y$ are summed up. However, in the bulk of the N region, the spin current is absent because the spin currents of electrons with opposed $k_x$ cancel each other. This compensation is disturbed when an electron current flows through the interface resulting in the net total spin current in the bulk.

The fact that this current exists in the N region far away from the interface, where no SOI is present, is caused by the infinite size of the interface scatterer in the $y$ direction. It is obvious, that the transverse spin currents should decay with distance deep into the N region because of scattering processes not considered here.

\begin{figure}
 \centerline{\includegraphics[width=8cm]{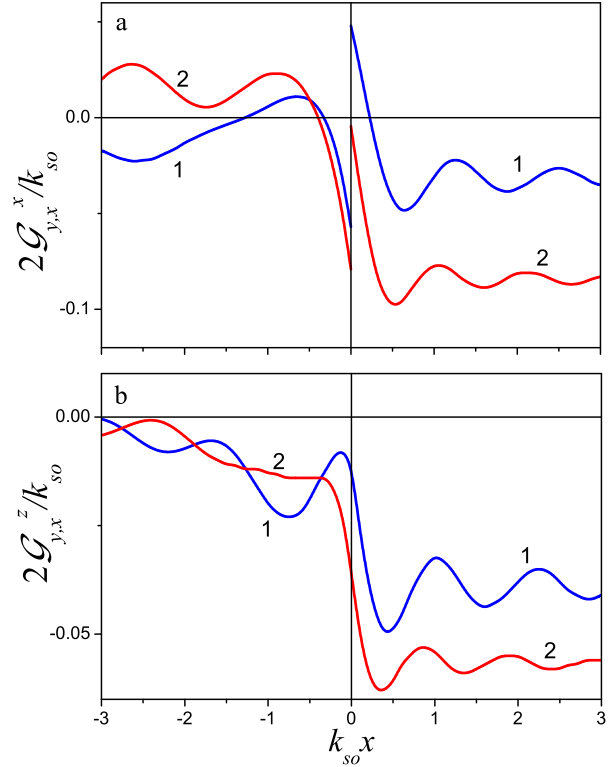}}
 \caption{Spatial distributions of $x$ and $z$ components of Hall spin conductivity (a) and (b) for different Fermi levels $\mu$ in the presence of a perpendicular current. Curve 1 -- $\mu=0.5E_{so}$,  curve 2 -- $\mu=2.5E_{so}$. $U_0=6E_{so}$ and $\beta=0.01$.}
\label{SH_cond}
\end{figure}

The spin currents $\mathcal{J}_{y,x}^x$ and $\mathcal{J}_{y,x}^z$ increase linearly with $\Delta\mu_x$ and therefore can be characterized by spin Hall conductivities $\mathcal{G}_{yx}^{x}$ and $\mathcal{G}_{yx}^{z}$. As distinct from the data of Fig.~\ref{SC_tang}, they describe only the non-equilibrium part of the spin currents. The spatial distributions of the spin Hall conductivities are shown in Fig.~\ref{SH_cond}.

It is worth emphasizing that the spin-Hall current in the bulk of the N region increases with the Fermi energy even if $\mu \gg E_{so}$ approximately as $\mu^{1/2}$, in contrast to the case for the normal spin current $\mathcal{J}_{x, x}^y$ and the spin density $S_x^y$. Numerically for the data of Fig.~\ref{SH_cond} the density of the transverse spin current is $\sim 0.45 (\hbar/2e)J$ at $\mu=0.5E_{so}$ and exceeds $0.6 (\hbar/2e)J$ at $\mu=2.5E_{so}$. Thus this effect is more robust and we believe that it could be attractive for experimental observation. The mechanism of increasing the transverse spin current with $\mu$ is explained taking into account that this current is created mainly by electrons moving nearly parallel to the interface with the velocity $\sim \hbar k_F/m$.

The transverse spin current in the N region could be investigated experimentally by measuring the spin density accumulated near the lateral edges of the sample. The density of spins near edge is estimated as follows
\begin{equation*}
n_s=\frac{2L_{sr}}{\hbar D_s}\mathcal{G}_{yx}^{x}\Delta\mu\,,
\end{equation*}
with $D_s$ the spin diffusion coefficient, $L_{sr}$ the spin relaxation length.
For $\mathcal{G}_{yx}^x=0.05k_{so}$, $k_{so}=10^{5}$~cm$^{-1}$, $L_{sr}=2\cdot{10}^{-3}$~cm, $D_s=10^2$~cm$^2$s$^{-1}$ and $\Delta\mu=10^{-5}$~eV one finds $n_{s}= 3\cdot{10}^{9}$cm$^{-2}= 30 \mu$m$^{-2}$. This value of spin density is readily measurable in the experiments~\cite{Kato,Crooker}.

\section{Conclusion}

We have found substantial spin effects in a normal 2D electron gas in contact with a 2D medium with SOI, which lies in the same plane. The spin effects are caused by the spin-dependent scattering of electrons from the interface.

In the thermodynamic equilibrium  state, the boundary scattering results in the appearance of an edge spin current flowing along the boundary in both SOI and N regions, with spin polarization being directed in the plane perpendicular to the boundary.

Under the non-equilibrium conditions caused by a ballistic electron current flowing in the system perpendicularly or tangentially to the interface, the following effects are developed. The tangential electron current gives rise to the formation of the edge spin density in both SOI and N regions. In addition it leads to the non-equilibrium edge spin currents that are proportional to the square of the electron current.

If the electron current flows normal to the interface, the main effect is the appearance of the non-equilibrium spin current. It is proportional to the charge current. The spin current has two components: longitudinal and transverse with respect to the charge current. The  longitudinal component is polarized parallel to the boundary. The transverse component is polarized in the plane perpendicular to the boundary. By its origin this component is the spin Hall current caused by the spin-dependent scattering of electrons from the interface. In view of possible experimental realizations, it is essential that the spin currents penetrate into the N region to a distance which is determined by scattering processes in the bulk and can really reach several microns. The spin-polarization degree of the currents is high enough (about $40-60\%$).

\subsection*{Acknowledgments}
This work was supported by Russian Foundation for Basic Research (project No 08-02-00777) and Russian Academy of Sciences (programs ``Quantum Nanostructures'' and ``Strongly Correlated Electrons in Solids and Structures'').

%\end{twocolumn}
\end{document}